\documentclass[11pt]{article}
\usepackage{authblk}
\usepackage{amsmath,amssymb,amsthm}
\usepackage{geometry}
\usepackage{color}
\usepackage{subcaption}
\usepackage{graphicx}
\usepackage{float}
\usepackage{authblk} % Required for the [1] author/affil syntax

\makeatletter
\renewcommand*{\@fnsymbol}[1]{\ensuremath{\ifcase#1\or \star \or \sharp \or \ddagger \or \mathsection \or \mathparagraph \or \| \or \star\star \or \sharp\sharp \else\@ctrerr\fi}}
\makeatother

\title{QCNN with Rough Path Signature Kernels}
\date{\today}

\author[1]{Leonardo Nogueira Falabella\thanks{leonardo.falabella@ensta.fr}}
\author[1]{Vasily Sazonov\thanks{vasily.sazonov@cea.fr}}
\affil[1]{Université Paris-Saclay, CEA, List, F-91120, Palaiseau, France}

\providecommand{\keywords}[1]{\textbf{Keywords:} #1}

\begin{document}
\maketitle

\begin{abstract}
Time series analysis plays a vital role across a wide range of scientific and engineering domains but poses substantial computational challenges. A major difficulty arises from the time reparameterization invariance of time series data, which complicates the extraction of meaningful temporal features. In this work, we address the problem of time series classification by exploring the application of quantum computation techniques. We propose a hybrid quantum–classical architecture that integrates recent advances in quantum neural networks with the mathematical framework of path signatures, mitigating the impact of time reparametrization invariance. The architecture employs feature layers that compute a signature kernel between pairs of input paths—consisting of a reference path and a target path for classification—using either classical or quantum variational linear solvers (VQLS). These feature layers are followed by a Quantum Convolutional Neural Network (QCNN) to perform downstream learning tasks. We evaluate several realizations of the proposed architecture, differing in QCNN configurations, on a binary classification task involving time series representations of handwritten digits. Our experiments demonstrate the potential advantages of implementing path signature kernel layers within quantum circuits and provide an analysis of the computational limitations associated with the VQLS component.
\end{abstract}


\newpage

\section{Introduction}

Time series data, representing sequential measurements over time, play a pivotal role across scientific and industrial domains: from financial market fluctuations and physiological signals such as heart rate or EEG/MEG brain recordings to meteorological patterns. The ability to analyze temporal sequences is essential for modern data-driven decision-making \cite{lim2021time, fawaz2019deep, hyndman2018forecasting}. Machine learning provides powerful tools for such analyses; however, these methods often face fundamental challenges arising from inherent symmetries in the data \cite{otto2025symmetry}.

A key difficulty lies in the invariance of sequential data to time reparametrization—its essential characteristics remain unchanged under arbitrary accelerations, decelerations, or nonlinear warpings of the time axis \cite{sakoe1978dynamic}. Standard models, which depend on the explicit parametrization of time, struggle to account for this invariance. Rough path theory offers an elegant mathematical framework to address this issue through the notion of the \textit{signature} \cite{chevyrev_primer_2025, varzaneh2023}. The signature transform converts a path into an infinite-dimensional feature vector of its iterated integrals, capturing its geometric and algebraic structure in a way that is inherently invariant to time reparametrization. Despite its advantages, computing truncated signature features remains computationally intensive, as the number of terms grows exponentially with the dimension of the path and the truncation order \cite{chevyrev_primer_2025, varzaneh2023}. Consequently, efficient computation of path signature features represents a significant bottleneck for large-scale applications.

Important steps towards efficient computations of signature features were taken by introducing the signature kernels defined as the inner product of the truncated \cite{kiraly2019kernels} and infinite signatures \cite{salvi_sign_2021}. As shown in \cite{salvi_sign_2021}, this kernel for infinite signatures is the unique solution to a well-defined second-order hyperbolic linear partial differential equation (PDE) - a Goursat problem. This enables practical computation of the kernel without explicitly evaluating the signatures themselves, instead relying on efficient numerical PDE solvers over a discretized grid. The classical finite-difference scheme for this PDE, however, produces a large, sparse system of linear equations whose solution scales polynomially with grid resolution. 

In the present work, we take the next step by incorporating the signature kernel in the hybrid quantum/quantum-classical variational architecture that is aimed at the time series classification problem.
The structure of the proposed architecture is presented in Figure \ref{fig: classification pipeline}. First, the feature layers evaluate a signature kernel between pairs of input paths using the classical routine to solve kernel PDE or a Variational Quantum Linear Solver (VQLS \cite{bravo-prieto_variational_2023}) to solve associated system of linear equations derived from the finite-difference discretization of kernel PDE. These feature layers are subsequently followed by Quantum Convolutional Neural Network (QCNN) layers \cite{Cong2019}, which perform downstream learning and classification tasks directly on the quantum-encoded features.

\begin{figure}[H]
    \centering
    \includegraphics[
        scale=0.4,
        trim={0cm 5.5cm 4.5cm 2cm},
        clip
    ]{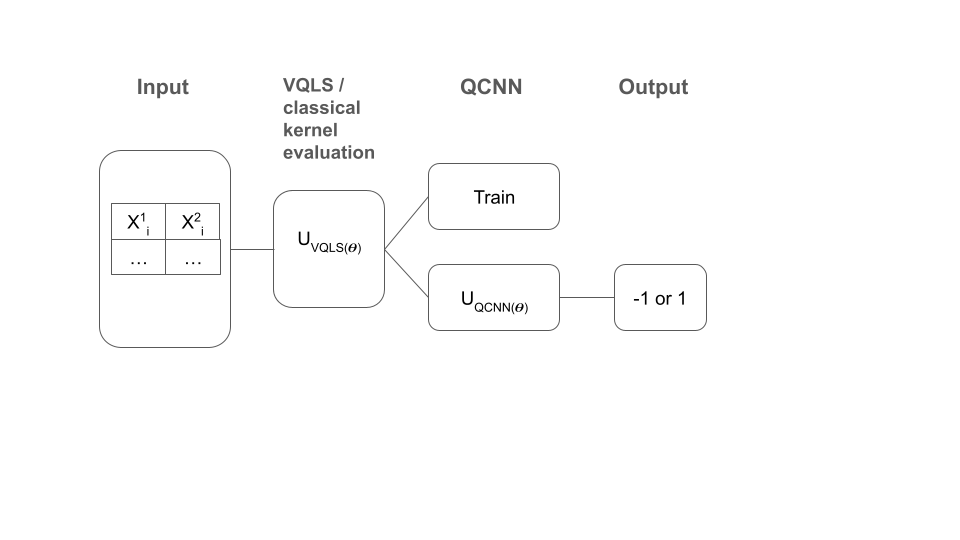}
    \caption{Schematic diagram for the hybrid quantum-classical classification pipeline.}
    \label{fig:classification_pipeline}
\end{figure}

We analyze computational limitations of the VQLS component of our architecture, and then, using classical routines for the signature kernel evaluations, we explore several realizations of the architecture based on different QCNN configurations. To demonstrate the framework’s potential, we apply it to a binary classification task involving handwritten digit recognition using the \textit{MNIST digits stroke sequence dataset} \cite{mnist-dataset}, where standard image data are transformed into time-ordered stroke sequences. This proof-of-concept application highlights both the advantages and current limitations of quantum-enhanced time series analysis, paving the way for future developments in quantum machine learning methods for sequential data.

The rest of the paper is organized as follows. In Section \ref{ker}, we introduce basic definitions of path signatures, describe the linear system of equations that represent the discretization of the PDE of the kernel. Section \ref{vqls} is dedicated to the solution of the discretized PDE solution with the VQLS method.
Then, in Section \ref{QCNN}, we present the QCNN architectures used in the current work. The last two sections are dedicated to applying the developed architectures to the digits from the MNIST dataset and to the concluding remarks.

\section{Signatures and kernels}
\label{ker}

A compact and highly descriptive representation of a time series (paths), useful for irregularly sampled multivariate sequential data applications, that summarizes geometric information about the path, can be achieved by the signature transform \cite{chevyrev_primer_2025}.
For a path $X: [a,b] \rightarrow \mathbb{R}^d$, we denote each of its coordinate paths by $(X^1, ..., X^d)$, where each $X^i:[a,b] \rightarrow \mathbb{R}$ is a real-valued path. The first-order terms of the signature are defined by
\begin{equation}
        S(X)_{a,b}^{i} = \int_{a<t<b} dX_t^{i} = X_b^i - X_a^i,
    \label{eq: first order sig terms}
\end{equation}
which simply gives the displacement of the coordinate $X^i$ in the path. The second-order terms are given by the double-iterated integral.
\begin{equation}
        S(X)_{a,b}^{i, j} = \int_{a<t<b} S(X)_{a,t}^{i} dX_t^{j} = \int_{a<s<t<b} dX_s^{i} dX_t^{j}.
        \label{eq: second order sig terms}
\end{equation}
We can continue recursively, and for the \textit{k-fold iterated integral} along the collection of indexes $i_1, ..., i_k \in \{1, ..., d\}$, with $k \geq 1$, we define
\begin{equation}
S(X)_{a,b}^{i_1,...,i_k} = \int_{a<s<b} S(X)_{a,s}^{i_1,...,i_{k=1}} dX_s^{i_k} = \int_{a<t_k<b} ... \int_{a<t_1<t_2} dX_{t_1}^{i_1} \ldots dX_{t_k}^{i_k}. 
\end{equation}
Then, the \textit{signature} of a path $X:[a,b] \rightarrow \mathbb{R}^d$, denoted by $S(X)_{a,b}$, is the infinite sequence of real numbers
\begin{equation}
    S(X)_{a,b} = \left(1, S(X)_{a,b}^1, ..., S(X)_{a,b}^d, S(X)_{a,b}^{1,1}, S(X)_{a,b}^{1,2}, ...\right),
\end{equation}
where, the "zeroth" order term is 1 by convention, and the superscript runs along the set of all possible \textit{multi-indexes} $\{(i_1, ..., i_k\, | \, k\geq 0, i_i, ..., i_k \in \{1, ..., d\}\}$.

It is not difficult to see, that being \textit{path integrals} of the type $\int_{a}^b Y_t\,dX_t = \int_{a}^b Y_t \dot{X_t}\,dt$, signatures are invariant under time reparametrizations $\psi : [a, b] \rightarrow[a,b]$, where $\psi$ is a subjective, continuous, non-decreasing function that maps the a time interval $[a, b]$ to itself. The signatures act as a filter that systematically removes the infinite-dimensional group of reparametrization symmetries.

The truncated signatures exhibit an exponential increase in the number of features, which limits their direct applicability to machine learning tasks in which the data streams reside in relatively low-dimensional ambient spaces. In data science, kernel methods have proven to be powerful learning techniques for high-dimensional (not necessarily sequential) inputs. Using the \textit{kernel trick}, many commonly used kernels can be computed efficiently without explicitly evaluating the corresponding feature maps. Here we will focus on the kernel computing the inner product of two infinite vectors of signatures. 
According to \cite{salvi_sign_2021}, the signature kernel is defined as follows. Consider two compact intervals $I = [u,u']$ and $J:[v, v']$ and paths defined over them $X : I \rightarrow \mathbb{R}^d$ and $Y : J \rightarrow \mathbb{R}^d$ and their corresponding signatures $S(X)_s$ and $S(Y)_t$, denoting the signature over the interval $[u, s]$ and $[v, t]$ for any $s \in I$ and $t \in J$, respectively. Then, the signature kernel $k_{X,Y}:I \times J \rightarrow \mathbb{R}$ is defined by the inner product of two (untruncated) signatures
\begin{equation}
    k_{X,Y}(s, t) = \langle S(X)_s, S(Y)_t \rangle,
\end{equation}
this kernel solves the linear, second-order, hyperbolic PDE
\begin{equation}
    \frac{\partial^2 k_{X,Y}}{\partial s \, \partial t} = \langle \dot{X}_s, \dot{Y}_t \rangle \cdot k_{X,Y}, \quad 
k_{X,Y}(u, \cdot) = k_{X,Y}(\cdot, v) = 1
\label{eq: signature kernel pde}
\end{equation}
where $\dot{X}_s = \left. \frac{dX_p}{dp} \right|_{p=s}, \quad \dot{Y}_t = \left. \frac{dY_q}{dq} \right|_{q=t}$ are the derivatives of $X$ and $Y$ at times $s$ and $t$ respectively. This equation is an example of a \textit{Goursat problem}, \cite{edouard_goursat_course_1916}, with a uniqueness solution on the bounded domain,
\begin{equation}
    D:= \{ (s,t) \,| \, u \leq s \leq u', v \leq t \leq  v' \} \subset I \times J \,.
\end{equation}
The discretization of \eqref{eq: signature kernel pde} is achieved in two steps: first, by writing the equation in its integral form and approximating the double integral,
\begin{equation}
\begin{split}
    k_{X,Y}(s,t) &\approx k_{X,Y}(s,v) + k_{X,Y}(u,t) - k_{X,Y}(u,v) \\
     & + \frac{1}{2} \langle \dot{X}_s, \dot Y_t\rangle (k_{X,Y}(s,v) + k_{X,Y}(u,t))(u-s)(s-v).
\label{PDE2}
\end{split}
\end{equation}
and then, by expressing \eqref{PDE2} on the grid defined by $D_I = \{ u=u_0 < u_1 < \ldots < u_{m-1} < u_m = u'\}$, a partition of the interval $I$, and $D_J = \{ v=v_0 < v_1< \ldots < v_{n-1} < v_n = v' \}$, a partition of the interval $J$. The \textit{finite difference scheme} on the grid $P_0 := D_I \times D_J$ and its dyadic refinements is given by
\begin{equation}
\begin{split}
    \hat k (s_{i+1}, t_{j+1})& =  \hat k (s_{i+1}, t_{j}) + \hat k (s_{i}, t_{j+1}) - \hat k (s_{i}, t_{j}) \\ &+ \frac{1}{2} \langle X_{s+1} - X_s, Y_{t+1} - Y_{t} \rangle (\hat k (s_{i+1}, t_{j}) + \hat k (s_{i}, t_{j+1})), \\ & \hat k (s_{0}, \cdot) = \hat k ( \cdot , t_{0}) = 1.  
\end{split}
\label{difscheme}
\end{equation}
This scheme has a time complexity of $O(d^2 2^{2\lambda}mn)$ on the grid $P_\lambda$, where $\lambda$ is the dyadic refinement order of the grid, $d$ is the dimension of the input streams $X, Y$ and $m,n$ denote their lengths. 
More sophisticated approximation schemes can be derived by applying higher-order quadrature techniques to approximate the double integral, as they are applied in the algorithm implementation of \textit{sigkernel} \cite{salvi_sign_2021}. Noticeably, this framework avoids the exponential complexity in the dimension and truncation level of signatures, since it does not require explicit computation of these objects.

\section{VQLS block}
\label{vqls}
To reformulate the evaluation of the signature kernel as a quantum routine, we rewrite the finite-difference scheme \eqref{difscheme} as a linear system of equations
\begin{equation}
    A\mathbf{x} = \mathbf{b}\,.
\label{LinSys}
\end{equation}
Let $(m + 1), (n + 1)$ be the number of points on the grid in the $s$ and $t$-directions, respectively. The equation for the discretized kernel values $\textbf{k} = [k_{1,1}, k_{1,2}, \ldots, k_{i, j}, \ldots, k_{m, n-1}, k_{m, n}]^T$ for $i=1, \ldots, m$ and $j = 1, \ldots, n$ (the values on the boundaries are known), can be rewritten as
\begin{equation}
    k_{i,j} = k_{i,j-1} + k_{i-1, j} - k_{i-1, j-1} + c_{i,j} (k_{i, j-1} + k_{i-1, j}), 
\label{PDE3}
\end{equation}
where $c_{i,j} = \frac{1}{2} \langle X_{s_i} - X_{s_{i-1}}, Y_{t_{j}} - Y_{t_{j-1}} \rangle$.
The following shows that \eqref{PDE3} written in the form \eqref{LinSys} contains the information about boundary conditions in the vector $\mathbf{b}$, and highlights the low triangle structure of the resulting linear system.
Consider the equation \eqref{PDE3} for each point on the grid: for $i=1, j=1$
\begin{equation}
k_{1,1}  = 1 + 1 -1 + c_{1,1}(1+1)
\implies k_{1,1} = 1+ 2c_{1,1},
\end{equation}
now for $i=1, j\geq 2$
\begin{equation}
    k_{1,j} = k_{1, j-1} + 1 - 1 + c_{1,j} (k_{i,j-1} - 1)
    \implies k_{i,j} - (1+c_{1,j})k_{1, j-1} = c_{1,j},
\end{equation}
for $i\geq2, j=1$
\begin{equation}
    k_{i, 1} = 1 + k_{i-1, 1} - 1 + c_{i,1}(1-k_{i-1,1})
    \implies k_{i,1} - (1+c_{i,1}) k_{i-1, 1} = c_{i,1},
\end{equation}
and finally for the interior points $i\geq2, j\geq2$
\begin{equation}
    k_{i,j} - (1+c_{i,j})k_{i,j-1} - (1+c_{i,j})k_{i-1, j} + k_{i-1,j-1} = 0.
\end{equation}

Ignoring the complexity of loading classical data to the quantum registers, the resulting system can be solved with an exponential speedup by the Harrow--Hassidim--Lloyd (HHL) algorithm \cite{harrow_quantum_2009} having the $O(\log N\, s\, \kappa^2/\epsilon)$ runtime - logarithmically with the system size $N$, quadratically in condition number, $\kappa$, and linearly in sparity $s$, offering exponential speedup over classical methods \cite{harrow_quantum_2009, childs_quantum_2017}. However, current NISQ devices cannot effectively implement HHL due to the required large circuit depth, limited qubit counts, noise accumulation, and costly state preparation \cite{marfany2024identifying, yalovetzky2021solving}. Therefore, for our architecture, we use the quantum variational algorithm (VQLS) \cite{bravo-prieto_variational_2023}, which encodes the solution in the parametrized quantum state prepared and measured on a quantum device and optimized with classical computations. Lower qubit count and short circuit depth, with rare need of fully connected qubits makes VQLS applicable on the NISQ devices. In addition, the variational core makes this algorithm more resistant to interactions of quantum computers with the surrounding media, due to the possibility of the Optimal Parameter Resiliance (OPR), where the optimal parameters remain unaffected by specific noise models \cite{bravo-prieto_variational_2023}. Despite some practical success of the VQLS algorithm, its rigorous complexity remains unknown. 

The VQLS algorithm aims to variationally prepare a quantum state that is proportional to the solution vector, $\bf{x}$, of the linear system \eqref{LinSys}. It assumes that there exists a decomposition of $A$ into a linear combination of $L$ unitaries of the form
\begin{equation}
    A = \sum_{l=1}^L c_l A_l,
\label{decomposition}
\end{equation}
where $A_l$ are unitaries and $c_l$ are complex numbers. Also, it is assumed that the condition number $\kappa < \infty$ and $||A|| \leq 1$, and that the $A_l$ can be efficiently implemented. 
In practice, one usually implements \eqref{decomposition} by decomposing matrix $A$ as a sum of Pauli strings, or two the sum of four unitaries, as
\begin{eqnarray}
    A &=& A_R + i A_I\,, \nonumber\\
    A_{R,I} &=& \frac{1}{2} \left[(A_{R,I} + i\sqrt{I - A_{R,I}^2}) + (A_{R,I} - i\sqrt{I - A_{R,I}^2}) \right]\,
\label{UA}
\end{eqnarray}
where $A_{R,I}$ are the Hermitian and anti-Hermitian parts of the matrix $A$, and matrices $(A_{R,I} \pm i\sqrt{I - A_{R,I}^2})$ are unitary. For our architecture, we use the latter variant \eqref{UA}. 

The VQLS algorithm receives the $A$ decomposition and an efficient gate sequence $U$ such that $U|0\rangle = |b\rangle$ as the inputs. 
Using the variational unitary $V(\theta)$, in our case given by the hardware-efficient ansatz with standard alternating layers of single-qubit $Y$ rotation and $CNOT$ entanglement gates, and the number of layers $L_{VQLS} = 4$,
\begin{equation}
    U_{VQLS}(\theta) = \prod_{l=1}^{L_{VQLS}} \left( \bigotimes_{i=1}^{n_b} R_y(\theta_{l,i}) \right) \cdot U_{\text{entangle}}\,,
\end{equation}
the algorithm prepares trial states $U_{VQLS}(\theta)|0\rangle = |x(\theta)\rangle$ optimized according to the cost functions. The minimization of cost functions leads to the minimization of the $A|x(\theta)\rangle$ component orthogonal to $|b\rangle$.
The ultimate set of parameters $\theta$ gives the final quantum state $U_{VQLS}(\theta^*) |0\rangle = |x(\theta^*)\rangle = \mathbf{x}/||\bf{x}||$. 
Following \cite{bravo-prieto_variational_2023} we use the global cost function
\begin{equation}
    C_G = \frac{Tr\left(|\psi\rangle \langle \psi | ( \mathbf{1} - |b\rangle \langle b|)\right)}{\langle \psi | \psi \rangle},
    \label{eq: local cost function}
\end{equation}
and the local one, designed to overcome the barren plateaus appearing in the global optimization,
\begin{equation}
    C_L = \frac{\langle x | H_L | x \rangle}{\langle \psi | \psi \rangle}.
\end{equation}
Here, the effective Hamiltonian $H_L$ is
\begin{equation}
    H_L = A^\dagger U \left( \mathbf{1} - \frac{1}{n}\sum_{j=1}^n |0_j\rangle \langle 0_j | \otimes \mathbf{1}_{\bar{j}} \right) U^\dagger A ,  
\end{equation}
with $n$ being the number of qubits used to represent the solution state $|x\rangle$, and $|0_j\rangle$ the zero state on qubit $j$, and $\mathbf{1}_{\bar{j}}$ the identity on all qubits except qubit $j$.

\section{Quantum Convolutional Neural Networks (QCNN) block}
\label{QCNN}

Quantum Neural Networks (QNNs) encompass a broad class of machine learning models constructed from parameterized quantum circuits (PQCs). Designed to harness quantum supremacy for complex computational workloads, QNNs have demonstrated significant promise across diverse applications, particularly in supervised classification tasks. Mirroring their classical counterparts, QNNs support various structural architectures. Since the layered architecture naturally allows the initial layers to compute kernels of rough path signatures, we restrict our attention in this work to the Quantum Convolutional Neural Network (QCNN) paradigm \cite{Cong2019}.

A QCNN typically employs a hierarchical, tree-like circuit architecture in which the number of active qubits is progressively reduced across successive layers. This reduction is analogous to the pooling operation in classical Convolutional Neural Networks (CNNs), enabling hierarchical feature extraction while reducing the dimensionality of the quantum state. The canonical QCNN architecture consists of alternating convolutional and pooling layers, followed by a final measurement stage. In the following, we describe the components of the QCNN architecture relevant to our study, based on the framework proposed in \cite{hur_quantum_2022}, together with several minor modifications introduced to accommodate information encoded through signature kernels.

\subsection{Quantum Data Encoding and Embeddings}

Before processing classical data within a quantum pipeline, it must be mapped from a classical feature space $\mathcal{X}$ into a high-dimensional quantum Hilbert space $\mathcal{H}$ via a quantum feature map $\phi: \mathcal{X} \rightarrow \mathcal{H}$, \cite{Rangaetal2024, Weigoldetal2021}. As evaluated by \cite{hur_quantum_2022} for embedding a classical data point $x \in \mathbb{C}^N$, the primary state-preparation approaches are amplitude encoding, which efficiently packs $2^n$ classical features directly into the amplitudes of n qubits, and angle or dense angle encoding, which embed classical features as rotation angles of single-qubit gates. In this work, we utilize amplitude encoding for the image representations of the signature kernels or images from the dataset, while we use angle encoding to include information about the signature kernel as a single number when loading into the QCNN architecture.

\subsection{Convolutional Layers and Ansatz Configurations}
The convolutional layers of a QCNN apply parameterized, local two-qubit unitary gates to adjacent qubit pairs to capture spatial and local correlations within the quantum data. To preserve structural regularity and reduce parameter complexity, these layers enforce \textit{translational invariance}, meaning that identical parameterized blocks are duplicated across a given layer. 

The expressivity and entangling capability of the QCNN depend heavily on the choice of the two-qubit ansatz $U(\theta)$. Following the architectural configurations analyzed by \cite{hur_quantum_2022}, several distinct unitaries—drawn from fundamental group theory and expressive PQC classifications \cite{Sim2019}—can be integrated as convolutional filters:
\begin{itemize}
    \item \textbf{$U_{\text{TTN}}$:} A minimal Tree Tensor Network ansatz consisting of basic single-qubit rotations and a coupling CNOT gate.
    \item \textbf{Heuristic Circuits ($U_5, U_6, B_9, U_{13}, U_{14}, U_{15}$):} Specialized, hardware-efficient ansatzes characterized by varying degrees of parameterization and entangling structures designed to balance expressibility against gate errors.
    \item \textbf{$U_{\text{SO}(4)}$ and $U_{\text{SU}(4)}$:} General group-theoretic unitaries. $U_{\text{SO}(4)}$ provides a special orthogonal transformation parameterized by 6 angles, while $U_{\text{SU}(4)}$ represents the most general, highly expressive two-qubit transformation utilizing 15 parameters \cite{hur_qcnn_github}.
\end{itemize}

\subsection{Pooling Layers}
Pooling layers reduce the dimensionality of the quantum system by mapping information from a group of qubits down to a reduced subset. This reduction is achieved by measuring a designated fraction of the qubits (e.g., every alternate qubit) and applying conditional single-qubit unitaries $U(\theta)$ to the remaining active qubits based on the classical measurement outcomes. 

\begin{figure}[htbp]
    \centering
    \includegraphics[scale=0.4, trim=0 2.5cm 0 0, clip]{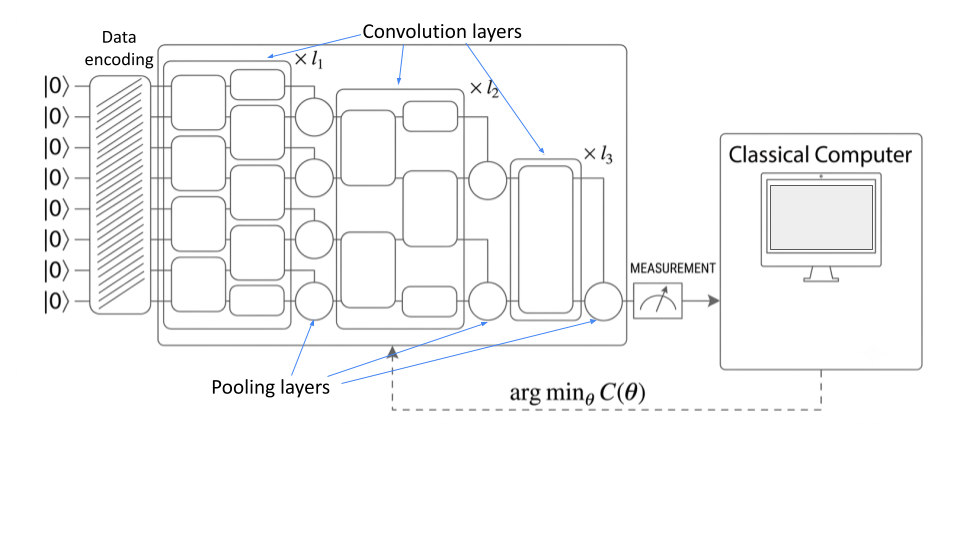}
    \caption{Schematic representation of the QCNN architecture applied to an 8-qubit system. The pipeline integrates a fixed quantum data encoding stage followed by three alternating layers of parameterized convolutional filters ($l_i$ denotes the number of filters in layer $i$) and pooling operations executed in a translationally invariant manner.}
    \label{fig: QCNN architecture}
\end{figure}

\subsection{Adapted Frameworks for Signature Kernels and Image Fusion}
In this work, we leverage the baseline architectures proposed by \cite{hur_quantum_2022, hur_qcnn_github} specifically for binary classification tasks, introducing two target modalities for processing sequential and structural data:
\begin{enumerate}
    \item \textbf{Pure Kernel Substitution Architecture:} Rather than utilizing standard pixel-intensity image data, we substitute the full signature kernel matrix (computed against a chosen reference sample $X_{\text{ref}}$) directly as the primary network input. This matrix effectively serves as an alternative structural ``image'' and is mapped onto a standard 8-qubit QCNN topology.
    \item \textbf{Auxiliary Qubit Fusion Architecture:} To concurrently handle multimodal data, we fuse the raw image dataset with sequence-derived kernel signatures. Here, the original image features are mapped onto a baseline 8-qubit register. Simultaneously, a single scalar representing the final, integrated signature kernel (the bottom-right cell of the kernel matrix reflecting full-path interaction) is encoded onto an additional, auxiliary 9th qubit using a parameterized $R_Y$ rotation. This auxiliary qubit is then systematically entangled and integrated with the main 8-qubit register according to the pre-existing hierarchical convolutional and pooling layout of the QCNN, see Fig. \ref{fig:qcnn aux qubit circ}.

\begin{figure}[htbp]
    \centering
    \includegraphics[width=0.7\textwidth]{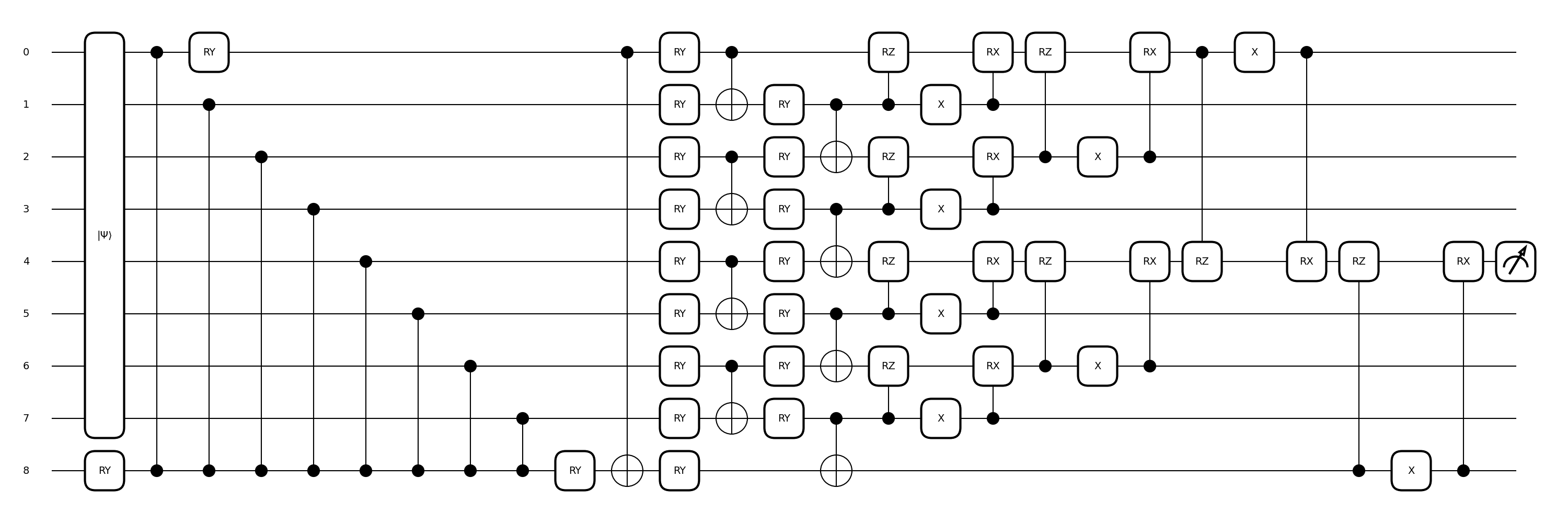}
    \caption{Circuit scheme of the extra qubit kernel architecture, exemplified with $R_Y$ and $CNOT$ gates as convolutional filters and parametrized pooling layers, corresponding to the architecture extended $U_{15}$, described above.}
    \label{fig:qcnn aux qubit circ}
    \end{figure}

\end{enumerate}

\subsection{Optimization}
Following the final pooling step, the network is reduced to a single output qubit. The expectation value of the Pauli-$Z$ operator on this remaining qubit constitutes the raw network prediction. 
The pipeline utilizes the Binary Cross-Entropy loss, 
\begin{equation}
    C_{\text{CE}}(\theta) = -\frac{1}{M} \sum_{i=1}^M \left[ y_i \log(p_i) + (1 - y_i) \log(1 - p_i) \right] \,,
\end{equation}
where $y_i \in \{1, 0\}$ are standard labels, $M$ is the number of training set samples, and the network's continuous prediction is converted to the probability $p_i$ of measuring the readout qubit in state $|0\rangle$, mapped via the transformation:
\begin{equation}
    p_i = \frac{\langle Z \rangle + 1}{2}\,.
\end{equation}

During training, a classical optimizer (e.g., Adam, COBYLA) updates the gate parameters $\theta$ to find an optimal set $\theta^* = \arg\min_\theta C(\theta)$. Once training is complete, the class label for an unseen data point $\tilde{x}$ is determined via $\tilde{y} = U(\tilde{x}, \theta^*)$, namely by application of the trained model.

A key advantage of the QCNN architecture is its highly efficient scaling behavior. Owing to the hierarchical reduction of qubits, the circuit depth scales as $\mathcal{O}(\log n)$ for an $n$-qubit input. Furthermore, because translational invariance dictates that parameters are shared within each layer, the total number of unique trainable parameters scales strictly as $\mathcal{O}(\log n)$. 
When computing gradients analytically on quantum hardware using the \textit{parameter-shift rule}, the total number of circuit executions per optimization step is directly proportional to the number of independent parameters.

\section{Application to the MNIST Dataset}
\label{sec:application_mnist}

In our experiments, we focus on handwritten digit classification framed as a binary $0$-vs-$1$ task, where the objective is to distinguish the digit $0$ from the digit $1$. For this purpose, we evaluate our models using two variants of the MNIST digit stroke sequence dataset. While the original MNIST dataset comprises $70{,}000$ grayscale, $28 \times 28$ images normalized to $[0,1]$ and partitioned into the canonical $60{,}000$ training and $10{,}000$ test samples, our Quantum Convolutional Neural Network (QCNN) utilizes only the subset containing digits $0$ and $1$. Specifically, this filtered dataset consists of $12{,}665$ training samples (comprising $5{,}923$ zeros and $6{,}742$ ones) and $2{,}115$ test samples (comprising $980$ zeros and $1{,}135$ ones), yielding a total of $14{,}780$ samples with an approximate split of $85.7\%$ training and $14.3\%$ testing data.

The end-to-end pipeline was constructed using a decoupled hybrid stack. All fundamental quantum circuits, state preparation workflows, and the Variational Quantum Linear Solver (VQLS) configurations were designed and executed using the \texttt{Qiskit} SDK \cite{javadi-abhari_qiskit_2024}. The Quantum Convolutional Neural Network (QCNN) architectures, parameter-sharing layers, and hybrid optimization routines were implemented via the \texttt{PennyLane} framework \cite{bergholm_pennylane_2022}. All numerical experiments were conducted on high-performance, noiseless statevector simulators to isolate algorithmic expressivity from environmental decoherence.

\subsection{Quantum versus Classical Kernel Evaluation}
\label{subsec:kernel_evaluation}

The mathematical evaluation of signature kernels can be formulated as solving a specialized second-order hyperbolic partial differential equation (PDE). Theoretically, this linear system can be accelerated on quantum hardware by mapping the discretized PDE to a quantum linear system problem ($Ax = b$). While the landmark HHL algorithm offers exponential speedups for such tasks, its deep coherence time requirements render it entirely impractical for Noisy Intermediate-Scale Quantum (NISQ) devices. 

To address this on near-term hardware, we developed and investigated an optimization routine using the Variational Quantum Linear Solver (VQLS) framework \cite{bravo-prieto_variational_2023}. Our formulation prioritizes a symmetric decomposition of the initial coefficient matrix $A$, which yields four distinct linear unitary terms ($A = \sum_n c_n A_n$) and demonstrates superior numerical stability in initial testing. To mitigate the impacts of barren plateaus and ensure algorithmic scalability, we utilize a local cost function $C_L(\boldsymbol{\theta})$ optimized via the gradient-free Constrained Optimization by Linear Approximation (COBYLA) algorithm.

However, scaling VQLS to accommodate realistic data trajectories introduces severe hardware challenges. Let a given handwriting sample be described by $T+1$ discrete time steps. The corresponding signature kernel matrix yields an un-symmetric system of $T^2$ unknowns. Applying a standard Hermitian embedding trick alongside any necessary zero-padding expands the final system dimension to at least $2T^2 \times 2T^2$. For typical MNIST stroke sequences containing between 20 and 50 time steps, the linear operator scales to a minimum size of $800 \times 800$, requiring an absolute minimum of 10 fully connected qubits.

\begin{figure}[htbp]
    \centering
    \begin{subfigure}{0.48\textwidth}
        \centering
        \includegraphics[width=\textwidth]{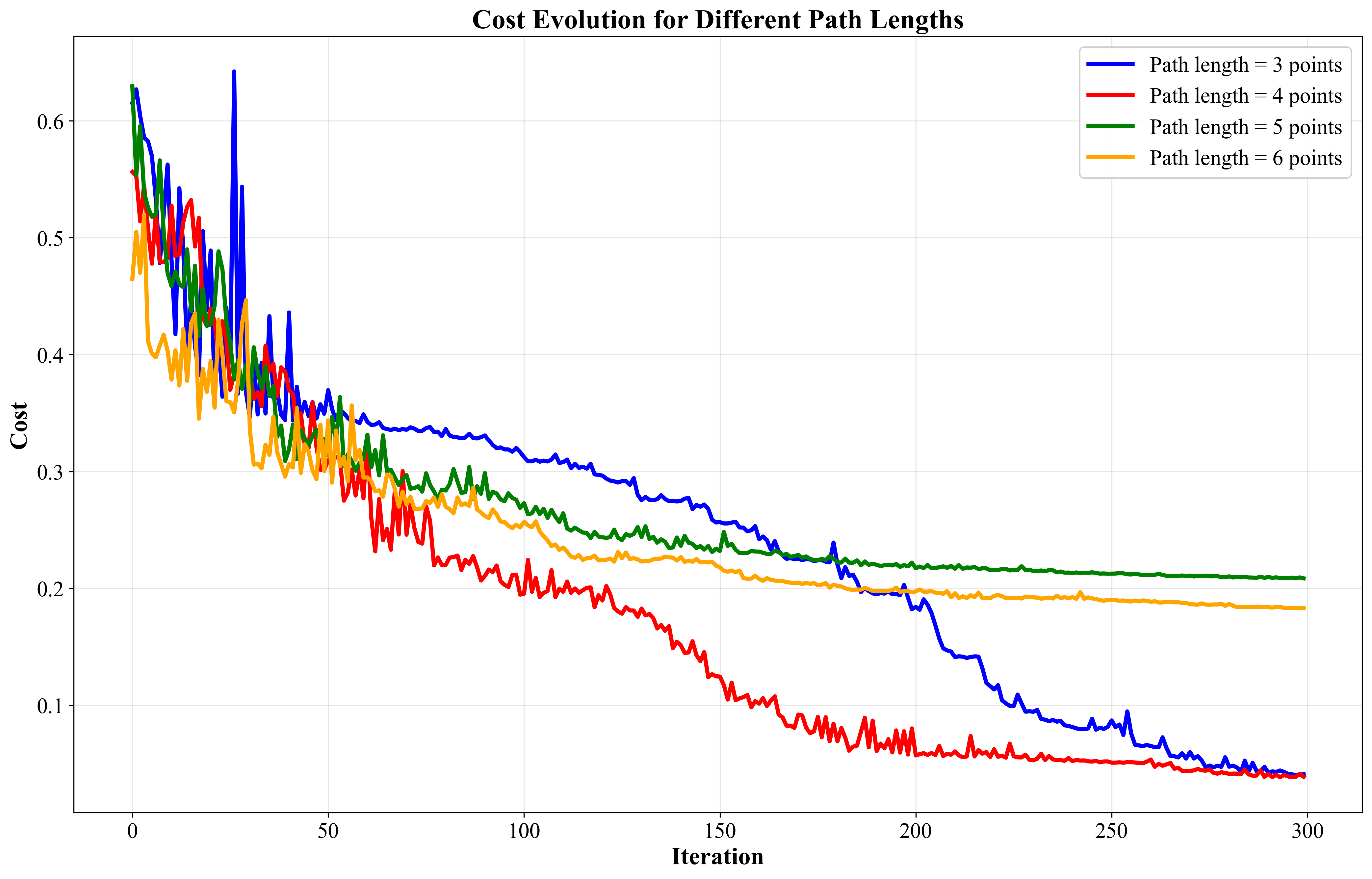}
        \caption{VQLS Cost Convergence}
        \label{fig:vqls_convergence}
    \end{subfigure}
    \hfill
    \begin{subfigure}{0.48\textwidth}
        \centering
        \includegraphics[width=\textwidth]{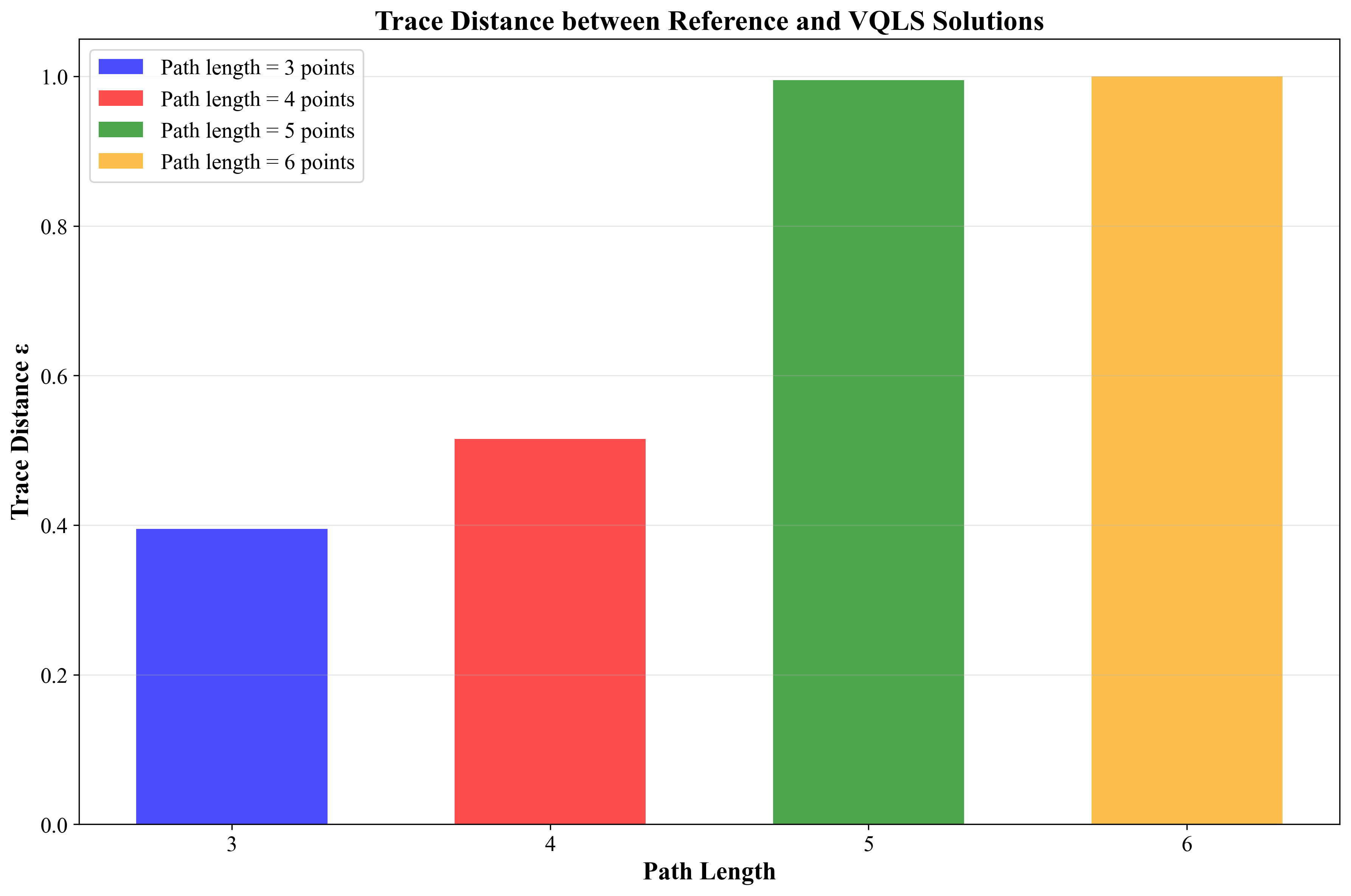}
        \caption{Overlap Fidelity Degradation}
        \label{fig:fidelity_degradation}
    \end{subfigure}
    \caption{Scalability benchmarks for the VQLS framework across varying path compression levels: (a) Optimization trajectories of the local cost function $C_L(\boldsymbol{\theta})$, and (b) The corresponding evolution of the overlap fidelity as a function of the system dimension.}
    \label{fig:vqls_benchmarks}
\end{figure}

To characterize these operational limits, we executed VQLS benchmarks across varying levels of path compression, as summarized in Figure \ref{fig:vqls_benchmarks}. For highly compressed paths (e.g., $3$–$4$ time steps), the cost function converges reliably, and the ansatz accurately reconstructs the true solution vector. However, as the path length increases to match realistic sample sizes (even for $5$-$6$ time steps), the optimization landscape becomes highly non-convex. This scaling behavior triggers a severe inflation of the matrix condition number $\kappa$, defined as the ratio of the maximum to minimum singular values of $A$. Theoretical bounds dictate that quantum linear solvers scale polynomially with $\kappa$ \cite{bravo-prieto_variational_2023}. Consequently, higher condition numbers cause the optimizer to trap in suboptimal local minima, leading to a marked degradation in the final overlap fidelity $\epsilon$, formally defined via the trace distance:
\begin{equation}
    \epsilon = \frac{1}{2}\text{Tr}\left| |x_0\rangle \langle x_0| - |x(\boldsymbol{\theta}_{\text{opt}})\rangle \langle x(\boldsymbol{\theta}_{\text{opt}})| \right|
    \label{eq:overlap_fidelity}
\end{equation}
where $|x_0 \rangle$ represents the exact classical solution and $|x(\boldsymbol{\theta}_{\text{opt}})\rangle$ is the variational state optimized under VQLS. 

Because a robust classification pipeline requires computing thousands of independent signature kernels across the training set, relying entirely on the variational quantum solver without severe lossy path compression proves unfeasible for near-term pipelines. For the subsequent classification benchmarks, we therefore bypass VQLS and execute the kernel evaluation classically utilizing the specialized \texttt{sigkernel} Python library \cite{salvi_sign_2021}. 

\begin{figure}[htbp]
    \centering
    \begin{subfigure}{0.5\textwidth}
        \centering
        \includegraphics[width=\textwidth]{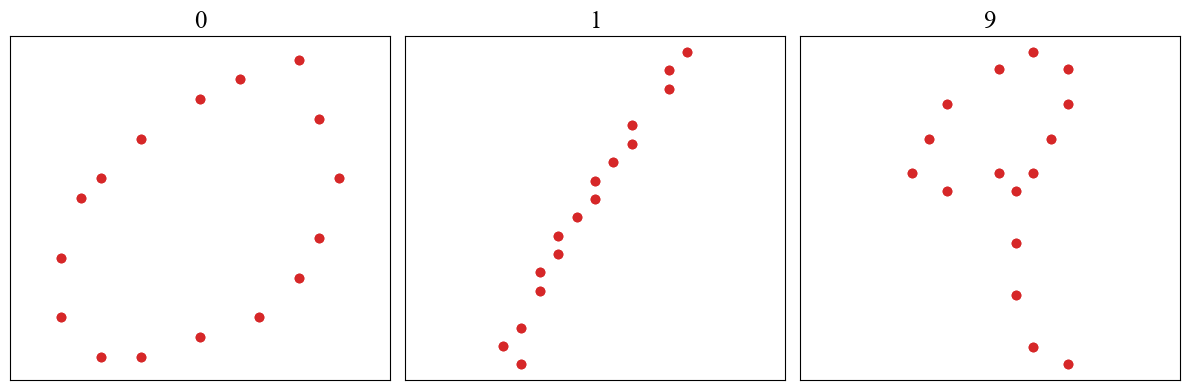}
        \caption{Sample Trajectories}
        \label{fig:mnist_samples}
    \end{subfigure}
    \hfill
    \begin{subfigure}{0.5\textwidth}
        \centering
        \includegraphics[width=\textwidth]{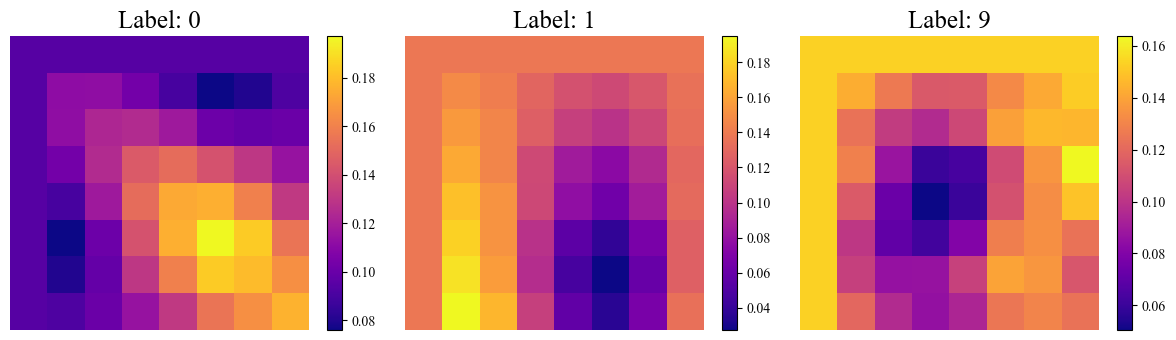}
        \caption{Kernel Matrices}
        \label{fig:kernel_matrices}
    \end{subfigure}
\caption{Visual analysis of the MNIST sequence dataset: (a) Characteristic downsampled pen-tip trajectories for the binary classification task, (b) Full signature kernel matrices generated by the SigKernel architecture, evaluated against a fixed reference sample $X_{\text{ref}}$ representing a digit $0$.}    \label{fig:mnist_comprehensive}
\end{figure}

However, to accommodate the hardware constraints of the subsequent Quantum Convolutional Neural Network (QCNN) architectures, where the input states are encoded into an $8$-qubit register, the original topological stroke sequences were downsampled. Specifically, the input paths were compressed to a fixed size of $16$ points, corresponding to $15$ time intervals. Each path is represented as a matrix of shape $(15, 3)$, which captures the $x$ and $y$ spatial coordinates alongside a binary indicator specifying whether the pen was down or up, while the temporal ordering of the points is implicitly preserved by their sequence in the data file. Examples of these downsampled $0$ and $1$ digits---as well as a digit $9$ to demonstrate an instance excluded from the binary $0$-vs-$1$ task---are illustrated in Fig.~\ref{fig:mnist_samples}. This mandatory downsampling inherently introduces a performance trade-off; restricting the coordinate sequence to so few points occasionally discards the fine geometric nuances required to robustly distinguish between certain digit classes, inevitably leading to instances of misclassification.

Both developed \textit{SigKernel}-based architectures consume these size-$(15, 3)$ paths, albeit through different integration mechanisms. The \textit{SigKernel} architecture compares two size-$(15, 3)$ paths to produce a full $15 \times 15$ kernel matrix, meaning the full image of the kernel is used. This matrix is subsequently flattened into a $225$-dimensional vector and loaded directly into the QCNN using Amplitude Embedding. Examples of these kernel matrices, evaluated against a fixed reference sample $X_{\text{ref}}$ representing a digit $0$, are visualized in Fig.~\ref{fig:kernel_matrices}. Alternatively, the hybrid \textit{SigKernel aux qubit} architecture compares the same size-$(15, 3)$ paths but utilizes only the main value of the kernel, reducing the evaluation to a single scalar value. This scalar is then encoded onto a dedicated auxiliary qubit, while the original $256$-pixel spatial image itself is uploaded directly into the remaining $8$ qubits unchanged.

\subsection{Performance Evaluation and Comparative Analysis}
\label{subsec:results_analysis}

Having resolved the kernel evaluation strategy, we analyze the performance of our hybrid QCNN implementations across the complete suite of convolutional ansatzes ($U_{\text{TTN}}$, $U_5$--$U_{15}$, $U_{\text{SO}(4)}$, and $U_{\text{SU}(4)}$). Performance characteristics are isolated by evaluating three distinct architectural configurations:
\begin{itemize}
    \item \textbf{\textit{Standard}:} The baseline spatial configuration where raw pixel inputs are encoded directly into an 8-qubit QCNN register.
    \item \textbf{\textit{SigKernel}:} The pure kernel substitution model, where spatial inputs are fully replaced by a truncated signature kernel matrix evaluated against a fixed reference trajectory $X_{\text{ref}}$, mapped across an 8-qubit register.
    \item \textbf{\textit{SigKernel aux qubit}:} The Auxiliary Qubit Fusion Architecture, which maps standard spatial image data across 8 qubits while concurrently encoding the total path interaction scalar (the bottom-right element of the kernel matrix) onto a 9th auxiliary qubit via a parameterized $R_Y$ rotation.
\end{itemize}

\begin{figure}[htbp]
    \centering
    \includegraphics[width=0.7\textwidth]{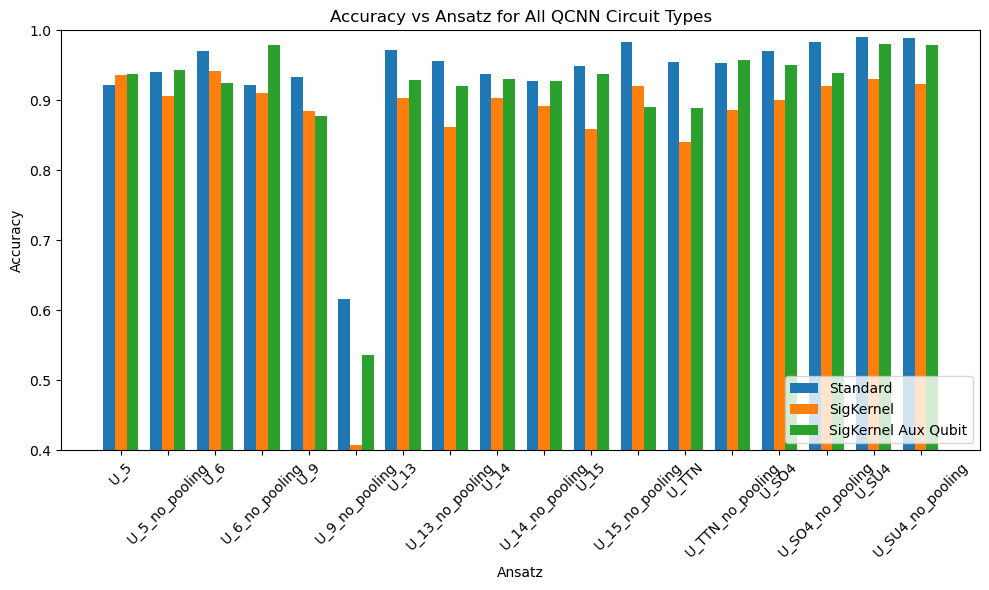}
    \caption{Binary classification accuracy for the $0$-vs-$1$ task benchmarks comparing the baseline \textit{Standard} image model, the pure \textit{SigKernel} substitution, and the multimodal \textit{SigKernel aux qubit} fusion architecture across various convolutional ansatzes.}
    \label{fig:accuracy_vs_circ_type_full_dataset}
\end{figure}

The empirical classification accuracies obtained for the $0$-vs-$1$ task across the various architectures are compiled in Fig. \ref{fig:accuracy_vs_circ_type_full_dataset}.

As illustrated in Fig.~\ref{fig:accuracy_vs_circ_type_full_dataset}, the introduction of signature kernel features yields a distinct performance enhancement for specific QCNN architectures. No clear dependence was observed regarding the number of tunable parameters within the QCNN component of the pipeline; instead, the magnitude of improvement appears to be primarily architecture-dependent.

Across the majority of the evaluated ansatzes, the {Auxiliary Qubit Fusion Architecture (\textit{SigKernel aux qubit})} consistently outperforms the pure kernel-based {SigKernel architecture}. This trend demonstrates that preserving the original spatial image structures while incorporating path signatures as an orthogonal, auxiliary quantum feature represents a highly effective integration strategy for parameter-efficient quantum classifiers. However, due to current constraints on the available number of qubits, simultaneously utilizing the full kernel matrix and the complete spatial image was not feasible. Scaling up the register size to accommodate both modalities concurrently could potentially yield uniform performance improvements across all evaluated architectures.

\section{Conclusions}
\label{sec:conclusions}
This research develops and evaluates a hybrid machine learning paradigm that merges the mathematical framework of rough path signatures with parameterized Quantum Convolutional Neural Networks (QCNNs). By re-framing sequence analysis through signature kernels, we established a method for transforming continuous, time-ordered trajectories into a compact, feature-rich algebraic representation compatible with quantum data-loading architectures.

A central focus of this work involved exploring the computation of signature kernels within a quantum framework via finite-difference discretization of the kernel's governing partial differential equation. While theoretically promising for end-to-end quantum acceleration, our evaluation of the Variational Quantum Linear Solver (VQLS) revealed severe scalability bottlenecks for near-term quantum devices. The rapid inflation of both system dimensions and matrix condition numbers restricts VQLS to heavily compressed sequences. Consequently, classical evaluation via libraries like sigkernel remains the only practical option for multi-sample training loops at present.

Despite these computational limitations, our empirical benchmarks on the MNIST dataset demonstrate that signature features can provide a meaningful structural prior for quantum neural networks in the classification domain. Although unconstrained QCNN circuits achieve performance comparable to standard image-based methods on raw data, the inclusion of signature kernel layers delivers accuracy improvements for some of the tested quantum architectures. 

Future research should investigate strategies to improve VQLS trainability on large-scale systems, such as employing QAOA-based ansatzes \cite{farhi2014quantum} or leveraging the inherent structure of the linear systems. Additionally, extending these hybrid architectures to multi-class classification and evaluating their behavior on real quantum hardware subject to noise and finite sampling will be critical next steps. Finally, applying this signature-based learning pipeline to complex datasets where conventional classical methods perform poorly may better isolate and reveal the true advantages of quantum-enhanced approaches.

\section{Acknowledgments}
This work has received support from the
French State managed by the National Research Agency
under the France 2030 program with reference ANR-22-
PNCQ-0002. We acknowledge the use of IBM Quantum
services for this work. The views expressed are those
of the authors, and do not reflect the official policy or position of IBM or the IBM Quantum team.
The authors acknowledge the use of the PennyLane library for implementing and simulating the quantum convolutional neural networks.

\section{Data availability}
The source codes that generated and processed the
data supporting the findings of this letter are available
from the authors upon reasonable request.

%\bibliographystyle{plain}
%\bibliography{qcnnrpt}

\begin{thebibliography}{26}

\bibitem{bergholm_pennylane_2022}
Ville Bergholm, Josh Izaac, Maria Schuld, et al. 
\newblock Pennylane: Automatic differentiation of hybrid quantum-classical computations. 
\newblock {\em arXiv preprint arXiv:1811.04968}, 2018.

\bibitem{bravo-prieto_variational_2023}
Carlos Bravo-Prieto, Ryan LaRose, M. Cerezo, Yigit Subasi, Lukasz Cincio, and Patrick J. Coles. 
\newblock Variational quantum linear solver. 
\newblock {\em Quantum}, 7:1188, 2023.

\bibitem{chevyrev_primer_2025}
Ilya Chevyrev and Andrey Kormilitzin. 
\newblock A primer on the signature method in machine learning. 
\newblock {\em arXiv preprint arXiv:1603.03788}, 2016.

\bibitem{childs_quantum_2017}
Andrew M. Childs, Robin Kothari, and Rolando D. Somma. 
\newblock Quantum algorithm for systems of linear equations with exponentially improved dependence on precision. 
\newblock {\em SIAM Journal on Computing}, 46(6):1920--1950, 2017.

\bibitem{Cong2019}
Iris Cong, Soonwon Choi, and Mikhail D.~Lukin. 
\newblock Quantum convolutional neural networks. 
\newblock {\em Nature Physics}, 15(12):1273--1278, 2019.

\bibitem{edouard_goursat_course_1916}
Edouard Goursat and Earle Raymond Hedrick. 
\newblock {\em A Course in Mathematical Analysis}. 
\newblock Ginn \& Company, 1916.

\bibitem{farhi2014quantum}
Edward Farhi, Jeffrey Goldstone, and Sam Gutmann. 
\newblock A quantum approximate optimization algorithm. 
\newblock {\em arXiv preprint arXiv:1411.4028}, 2014.

\bibitem{fawaz2019deep}
Hassan Ismail Fawaz, Germain Forestier, Jonathan Weber, Lhassane Idoumghar, and Pierre-Alain Muller. 
\newblock Deep learning for time series classification: a review. 
\newblock {\em Data Mining and Knowledge Discovery}, 33(4):917--963, 2019.

\bibitem{harrow_quantum_2009}
Aram W. Harrow, Avinatan Hassidim, and Seth Lloyd. 
\newblock Quantum algorithm for linear systems of equations. 
\newblock {\em Physical Review Letters}, 103(15):150502, 2009.

\bibitem{hur_quantum_2022}
Tak Hur, Leeseok Kim, and Daniel K. Park. 
\newblock Quantum convolutional neural network for classical data classification. 
\newblock {\em Quantum Machine Intelligence}, 4(1):3, 2022.

\bibitem{hur_qcnn_github}
Tak Hur, Leeseok Kim, and Daniel K. Park. 
\newblock Quantum convolutional neural network (qcnn), 2021. 
\newblock {\it{https://github.com/takh04/QCNN}}. Implementation of quantum convolutional neural networks for classical data classification.

\bibitem{hyndman2018forecasting}
Rob J.~Hyndman and George Athanasopoulos. 
\newblock {\em Forecasting: principles and practice}. 
\newblock OTexts, 2nd edition, 2018.

\bibitem{javadi-abhari_qiskit_2024}
Ali Javadi-Abhari, Matthew Treinish, Kevin Krsulich, Christopher J. Wood, Jake Lishman, Julien Gacon, Simon Martiel, Paul D. Nation, Lev S. Bishop, Andrew W. Cross, Blake R. Johnson, and Jay M. Gambetta. 
\newblock Quantum computing with qiskit. 
\newblock {\em arXiv preprint arXiv:2405.08810}, 2024.

\bibitem{kiraly2019kernels}
Franz J. Kiraly and Harald Oberhauser. 
\newblock Kernels for Sequentially Ordered Data. 
\newblock {\em Journal of Machine Learning Research}, 20(31):1--45, 2019.

\bibitem{mnist-dataset}
Yann LeCun and Corinna Cortes.
MNIST handwritten digit database. 
http://yann.lecun.com/exdb/mnist/, 2010.

\bibitem{lim2021time}
Bryan Lim and Stefan Zohren. 
\newblock Time-series forecasting with deep learning: a survey. 
\newblock {\em Philosophical Transactions of the Royal Society A}, 379(2194):20200209, 2021.

\bibitem{marfany2024identifying}
Giulia Marfany, Ievgeniia Sakhnenko, and Rainer Lorenz. 
\newblock Identifying bottlenecks of NISQ-friendly HHL algorithms. 
\newblock {\em arXiv preprint arXiv:2406.06288}, 2024.

\bibitem{otto2025symmetry}
Samuel E. Otto, Nicholas Zolman, J. Nathan Kutz, and Steven L. Brunton. 
\newblock A unified framework to enforce, discover, and promote symmetry in machine learning. 
\newblock {\em Journal of Machine Learning Research}, 26(248):1--83, 2025.

\bibitem{Rangaetal2024}
Deepak Ranga, Aryan Rana, Sunil Prajapat, Pankaj Kumar, Kranti Kumar, and Athanasios V. Vasilakos. 
\newblock Quantum machine learning: Exploring the role of data encoding techniques, challenges, and future directions. 
\newblock {\em Mathematics}, 12:3318, 2024.

\bibitem{sakoe1978dynamic}
Hiroaki Sakoe and Seibi Chiba. 
\newblock Dynamic programming algorithm optimization for spoken word recognition. 
\newblock {\em IEEE Transactions on Acoustics, Speech, and Signal Processing}, 26(1):43--49, 1978.

\bibitem{salvi_sign_2021}
Cristopher Salvi, Thomas Cass, James Foster, Terry Lyons, and Weixin Yang. 
\newblock The Signature Kernel is the solution of a Goursat PDE. 
\newblock {\em SIAM Journal on Mathematics of Data Science}, 3(3):873--899, 2021.


\bibitem{Sim2019}
Sukin Sim, Peter D. Johnson, and Al\'an Aspuru-Guzik. 
\newblock Expressibility and capability of parameterized quantum circuits for hybrid quantum-classical algorithms. 
\newblock {\em Advanced Quantum Technologies}, 2(12):1900070, 2019.

\bibitem{varzaneh2023}
Mazyar Ghani Varzaneh and Sebastian Riedel. 
\newblock Introduction to rough paths theory. 
\newblock {\em arXiv preprint arXiv:2302.04653}, 2023.

\bibitem{Weigoldetal2021}
Manuela Weigold, Johanna Barzen, Frank Leymann, and Marie Salm. 
\newblock Expanding data encoding patterns for quantum algorithms. 
\newblock {\em 2021 IEEE 18th International Conference on Software Architecture Companion (ICSA-C)}, 2021.

\bibitem{yalovetzky2021solving}
Rodrigo Yalovetzky, Patrick Minssen, Daniel Herman, and Marco Pistoia. 
\newblock Solving linear systems on quantum hardware with hybrid HHL++. 
\newblock {\em arXiv preprint arXiv:2110.15958}, 2021.

\end{thebibliography}

\end{document}